# The Rise of Open Science: Tracking the Evolution and Perceived Value of Data and Methods Link-Sharing Practices

**Hancheng Cao**[a,1], **Jesse Dodge**[c], **Kyle Lo**[c], **Daniel A. McFarland**[a], **and Lucy Lu Wang**[b,c,1]

[a]Stanford University; [b]University of Washington; [c]Allen Institute for AI; [1]Correspondence should be addressed to: hanchcao@stanford.edu, lucylw@uw.edu

**In recent years, funding agencies and journals increasingly advocate for open science practices (e.g. data and method sharing) to improve the transparency, access, and reproducibility of science. However, quantifying these practices at scale has proven difficult. In this work, we leverage a large-scale dataset of 1.1M papers from arXiv that are representative of the fields of physics, math, and computer science to analyze the adoption of data and method link-sharing practices over time and their impact on article reception. To identify links to data and methods, we train a neural text classification model to automatically classify URL types based on contextual mentions in papers. We find evidence that the practice of link-sharing to methods and data is spreading as more papers include such URLs over time. Reproducibility efforts may also be spreading because the same links are being increasingly reused across papers (especially in computer science); and these links are increasingly concentrated within fewer web domains (e.g. Github) over time. Lastly, articles that share data and method links receive increased recognition in terms of citation count, with a stronger effect when the shared links are active (rather than defunct). Together, these findings demonstrate the increased spread and perceived value of data and method sharing practices in open science.**

## Introduction

The Open Science movement is motivated by the desire to accelerate scientific discovery and enhance the efficiency of science (1) through increased access to scientific output (2) and the dissemination of public scientific resources (1, 3). Stakeholders have actively advocated for open science practices, such as the open sharing of data and method implementations to improve the science reproducibility and extensibility. For instance, the European Commission proposed steps including the necessary change of researcher rewards and incentives towards sharing knowledge, publishing, and outreach (4). The Biden-Harris Administration announced "New Actions to Advance Open and Equitable Research,"[*] including policies and funding opportunities to facilitate open science practices. More recently, the US National Science Foundation called for "policies and investments to make reproducible and replicable science easier for scientific communities to understand and execute and to embed reproducibility and replicability within the fundamental scientific method."[†] Some publication venues also recommend such practices as part of their publication guidelines, with research communities organizing around events such as the Machine Learning Reproducibility Challenge.[‡]

The sharing of data and methods are two important open science practices emphasized in prior work. Sharing data and methods enables better research reproducibility, fosters scientific collaboration by involving researchers from a broader range of disciplines (5, 6), and provides open benchmarks to quantify progress towards shared academic goals (7). While sharing data and methods has benefits, there are many barriers to adopting such practices, including personal attitudes towards open science (8), lack of time, funding, and/or institutional support (9, 10), as well as inadequate archival standards and platforms (11). As a result, insufficient numbers of scientific publications share their data and methods, e.g., prior work making direct data requests to authors were successful in only 27-59% of cases depending on the field and/or publication venue (12). Even papers that include a data availability statement (such as "data is available upon request") may fail to provide data when requested; in a 2018 study, researchers contacted the authors of 204 papers with such statements, and only 44% provided the requested data (13). Other works in this vein report similarly low proportions of data being made available by authors (14, 15).

On the other hand, a number of studies have evaluated the effects of data and methods sharing in specific journals and research communities, mostly through studying a small sample of manually curated papers, a specific publication venue, or publications that include data availability statements. These prior attempts have concluded that data sharing is associated with increased citation rates in specific contexts (16, 17); for instance, an examination of 85 cancer microarray clinical trial publications found that 48% of trials with publicly available microarray data received 85% of the aggregate citations (18). A larger-scale analysis of 532k journal articles published by Public Library of Science (PLOS) and BioMed Central (BMC) found that articles including statements linking to data repositories (around 20% of papers) have up to 25% higher citation impact (19); however, this work was limited to two publishers and only inspected the existence of data availability statements provided

[*] https://www.whitehouse.gov/ostp/news-updates/2023/01/11/fact-sheet-biden-harris-administration-announces-new-actions-to-advance-open-and-equitable-research/
[†] https://beta.nsf.gov/funding/opportunities/reproducibility-replicability-science
[‡] https://paperswithcode.com/rc2022

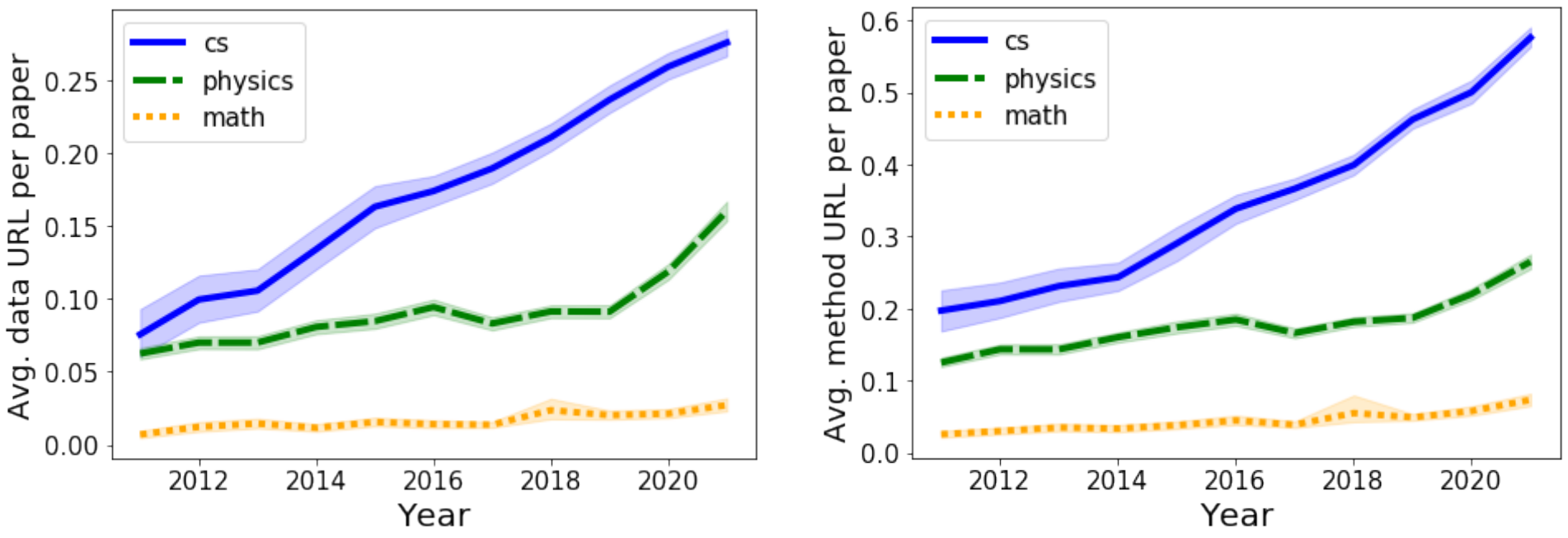

(a) Average data URLs per paper over time. (b) Average method URLs per paper over time.

**Fig. 1.** There is increased sharing of data and methods links in papers over time in the three fields, especially in Computer Science.

by authors, which may not cover other forms of data sharing and reuse in full text, and does not investigate whether data is actually retrievable at the specified locations.

Similarly, the practice of sharing methods (e.g., open code repository or tool library) has been found to increase understanding and reuse of work (20–22). Prior work has argued for the importance of making source code for computational research publicly available (23–26) and recommended standard steps for methods sharing (20). However, there are few empirical investigations into methods sharing practices in science and they are generally restricted to specific venues, e.g., Vandewalle (27) found that around 10% of papers from IEEE Transactions on Image Processing 2004-2006 contain references to code, and that this rate increased to about 25% for the same journal in 2017 (28).

As such, despite extensive literature on open science and data and methods sharing, several key issues persist with existing work: research on this topic has mostly used small-scale data within a limited time span and/or restricted to a few select (and often prestigious) venues (13, 19) or a specific research field of interest (7, 18). Prior analysis has also focused on data or methods sharing through the limited channels of data availability statements (19), whereas few publication venues require such statements, and researchers are increasingly likely to share data and methods through alternate channels, such as hosting these materials in online repositories and providing a web link in the paper text. Preliminary evidence (29) has shown that this type of practice leads to higher availability of resources when compared to having only a data availability statement. However, no large-scale analyses exist to assess broad patterns of researcher adoption of data and methods link sharing practices over time, i.e., it is unclear how prevalent link sharing practices are in science overall, whether these practices diffuse over time, and whether they indeed result in downstream citation benefits.

We address these limitations by performing a large-scale study of data and methods link-sharing practices in the fields of computer science, physics, and mathematics across 1.06 million scientific articles and preprints from arXiv from 2011 and 2021 using their full text data. We argue web links that host data and methods are an alternative, and likely more universal, means to study data and methods sharing at-scale that complements prior work using data availability statements. Through full text mining of web links, we are able to capture not only data and methods shared in dedicated data and methods availability sections, but also artifacts mentioned throughout these papers. Specifically, to identify links to data and methods, we extract URLs from the document source and train a natural language processing model to automatically classify each URL's type based on its contextual mention in the paper. This provides a scalable way to estimate a 'lower bound' of data and methods sharing through URL mentions in academic papers. We investigate how the practice of data and methods link sharing has changed over time in three academic fields, i.e., how data and methods links have been mentioned and reused over time. Moreover, we study how the loss of data and method links over time to link rot may impact a publication. Prior work has found that a large proportion of URLs in scientific papers stop working or are no longer retrievable after some time (this phenomenon is known as *link rot*) (30, 31). To investigate this issue, we perform HTTP requests to resolve data and methods links and estimate their retrievability. We find that the presence of data and methods links, in combination with their retrievable status, are predictive of a paper's citation impact.

## Methods

**Curating a URL dataset from arXiv.** In this study, we leverage data from arXiv, a popular web repository for hosting academic papers and preprints. ArXiv represents an important part of scientific literature, especially for the fields of computer science, physics, and mathematics, as researchers increasingly upload initial versions of their papers as preprints on arXiv before publication for more timely scholarly communication.

We retrieved all papers from arXiv using the bulk data access API [§]. Furthermore, we retrieved the following metadata associated with each arXiv paper from the Semantic Scholar database (32), including (1) the paper's citation count as of Oct 3, 2022 and (2) the arXiv subject fields[¶] associated with each paper. The highest level subject field classifications include computer science, math, physics, and other less well-represented subjects such as statistics and quantitative biology that we omit from our analysis. If a paper is associated with multiple subjects, we chose the primary subject classification to represent the subject of the paper. We also limit our analysis to publications uploaded to arXiv between 2011–2021 due to the relative sparsity of arXiv data before 2011.

§ https://arxiv.org/help/bulk_data
¶ https://arxiv.org/category_taxonomy

We extract all URL links from the full text of these papers. Then, we train a natural language processing model that classifies each link as resolving to a data, method, or other-type artifact, based on the context of the link's mention. As such, our approach offers a more universal approach to study data and methods sharing, one that does not rely on manual inspection (13) and which goes beyond publications with data or methods availability sections (19). We describe our approach in detail below.

***Retrieving links and associated context from arXiv LaTeX.*** To retrieve URL links in papers, we need to first parse the raw full text of papers in different formats into plain text, where arXiv papers are typically available in LaTeX or PDF formats. We kept arXiv papers (1,062,586 papers) with LaTeX source, which represent 88.0% of all papers on arXiv (1,208,034 papers). These papers are the focus of our study, primarily because parsing Latex files would introduce much fewer false positives and false negatives compared with parsing PDF files. We first leveraged the S2ORC-doc2json library[‖] (33) to parse LaTeX full text of paper into JSON format. We then use regular expressions to detect URLs from the JSON representation, retaining the URL, its mention context (the sentence where the detected links appear), as well as the corresponding section in which the URL was found. We further cleaned all extracted URLs through additional steps such as URL normalization to ensure its quality. Further details on URL extraction and normalization are presented in Appendix A.

***Link classification and validation.*** While links in papers could point to important contents such as data and methods, a significant proportion of links in papers are 'supplement' links, e.g. personal website, background citation like a piece of news, which are not central to research and open science practices. To distinguish links related to open science practices such as data and methods sharing from other less important links, we trained a classifier to determine the types of links to artifacts based on the link context. Specifically, we leverage the dataset released by Zhao et al. (34), which includes 3,088 manually annotated URLs and their contexts, with each artifact labeled into one of three categories: data, methods (tools, algorithms, etc.), and supplements [**].

We fine-tuned a SciBERT language model (35) on the task of URL classification using context (e.g. the authors write "we release our code at [URL]"; this [URL] would be classified as a methods link). As such, we are able to differentiate links more central to the research work from "supplementary" links (e.g., links to news articles motivating the research).

A 10-fold cross validation shows that our fine-tuned artifact classification model achieved 0.83 macro-F1, 0.83 macro-precision, 0.83 macro-recall, and 0.83 accuracy, suggesting high performance in using context to classify the type of link to artifact into material (data), methods (tools, algorithms, etc.), and supplements. To further validate the quality of URL artifact classification, we randomly sampled 100 extracted URL mentions from our arXiv dataset, and manually coded the URL linked artifact as data, methods or supplement based on the context of each link mention. When predicting over this validation set, our model achieves 0.94 macro-F1, 0.93 macro-precision, 0.97 macro-recall, and 0.95 accuracy, which demonstrates that the domain shift between the training dataset and our dataset did not lead to a drop in performance. We subsequently ran the final trained classifier on all extracted links and their contexts in our dataset.

***HTTP request to determine retrievability of links.*** Not all links are retrievable as some links will be no longer accessible and lost over time (30). This is reflected by the HTTP status code after requesting specific URLs. To determine the status of the artifact (e.g., whether the link is retrievable or not) so that we could analyze how artifact retrievability influence paper reception, we ran HTTP requests on the extracted and normalized links (we try pre-pending both `http://` and `https://` to the URL, and record the more successful status code), specifying a timeout threshold of 120 seconds, and a 6 second wait time between consecutive queries at the same domain (we tested multiple timeout and domain wait time thresholds, and selected values where the majority of requests are completed while the total time for request completion is within reason). We record the status code of each URL (or the status code of the final destination if the URL contains redirects). Finally, we consider the link as *alive* if it returns a 200 HTTP code and *problematic* if it returns a non-200 code, such as a 404, 403, or 503. All requests were made in late August, 2022. Additional results on link retrievability are shown in Appendix E.

***Dataset summary statistics.*** The final data consists of 1,062,586 arXiv papers in LaTeX format from 2011 to 2021 in the fields of computer science, physics, and mathematics, of which 199,102 (18.7% of sample) mention 289,875 unique URLs a total of 598,577 times (3.04 URLs listed per paper sharing URLs). Of all link mentions, 19.9% are data link mentions, 39.8% are method link mentions and 40.3% are supplementary link mentions. In total, there are 56,132 unique data links, 124,621 unique method links, and 143,918 unique supplementary links[††].After making HTTP requests on data and methods links, 83.1% of data links remained alive, while 86.6% of methods links remained alive. Detailed statistics are provided in Table 1, including splits by field and link type.

**Table 1. Statistics of the final analyzed data. We show the total number of papers and number of papers with links, split by field and by type of link (data, methods, or supplement). Unique counts are provided in parentheses.**

| | Papers | | Number of links (Unique link count) | | |
|---|---|---|---|---|---|
| | w/ Links | Total | Data | Method | Supplementary |
| **Computer Science** | 86,705 | 250,739 | 60,244 (35,995) | 119,751 (78,638) | 83,413 (63,232) |
| **Physics** | 92,948 | 533,364 | 53,662 (20,153) | 104,718 (42,521) | 110,693 (51,897) |
| **Mathematics** | 19,449 | 278,483 | 5,039 (3,879) | 13,719 (10,637) | 47,338 (35,372) |
| **Total** | 199,102 | 1,062,586 | 118,945 (56,132) | 238,188 (124,621) | 241,444 (143,918) |

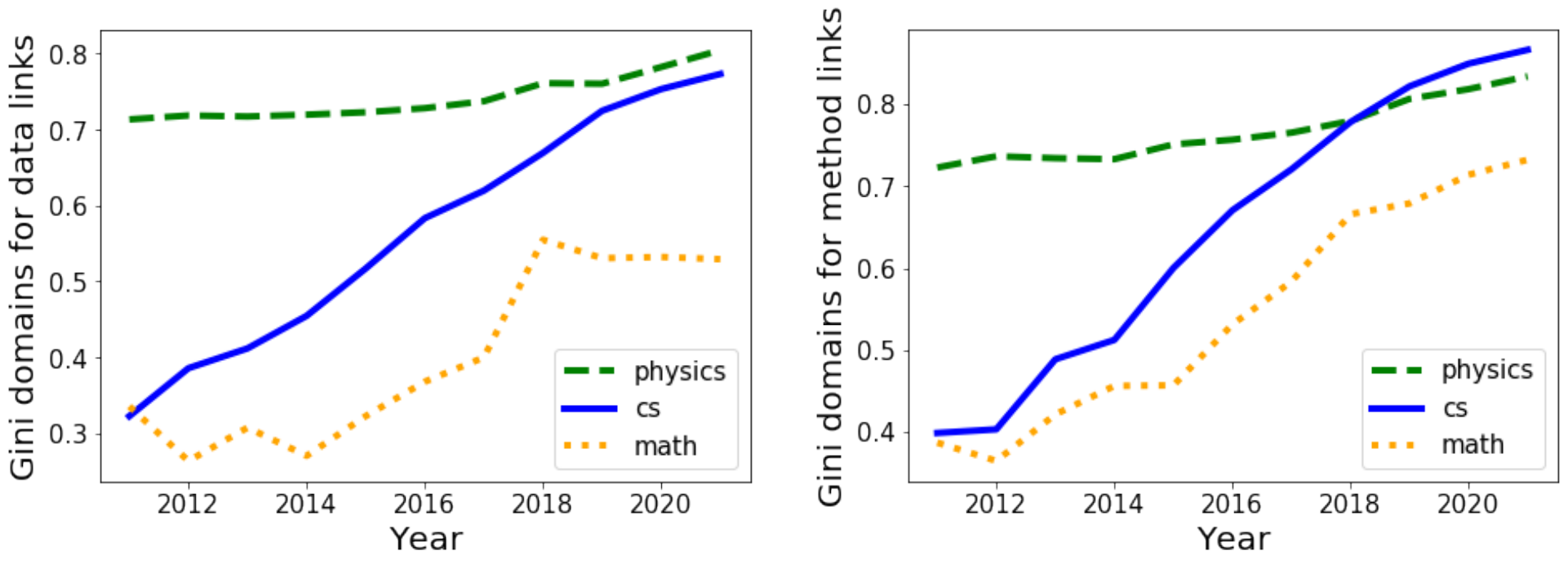

(a) Gini coefficient of domain in data URLs (b) Gini coefficient of domain in method URLs

**Fig. 2.** Gini coefficients show that data and methods links are becoming increasingly concentrated in popular web domains over time. This phenomenon can be observed in all three fields of study.

## Results

**Data and methods link sharing is increasing.** First, we find there has been growing usage of data and methods URLs in research papers over the years (Fig 1(a) and Fig 1(b)), indicating increasing prevalence of data and methods link sharing practices. For instance, the average computer science paper mentioned 0.08 data links in 2011 and up to 0.26 data links in 2020. Methods links were shared at higher rates, but also show increasing usage. Computer science papers mentioned on average 0.20 methods links in 2011 and up to 0.50 methods links in 2020. Similar growth trends can be observed in physics and math as well. These results suggest that over arXiv, authors from different fields are increasingly adopting data sharing and methods sharing practices. These trends suggest increasing adoption of open science research practices across fields, with the practice being especially strong in computer science.

Meanwhile, we also observe that an increasing proportion of all links in papers are to data, as shown in Fig 11(a). Around 15% of all links refer to data in 2011 within computer science while the percentage rose to 22% more recently. The proportion of data links in physics has also grown over time. Meanwhile, the proportion of methods links in the three fields remained relatively stable. Among the fields, math has the lowest proportion of data and methods links, and has the lowest growth rate for data and methods link usage compared with physics and computer science. This is unsurprising, since typical contributions in math do not involve software and the field is generally not reliant on empirical data.

**Data and methods links are increasingly reused over time.** The increasing use of data and methods links suggests that scholars are increasingly practicing scientific transparency. The growing *reuse* of data and methods links suggests something further—that scholars increasingly rely on research resources by others or practice reproducibility. We find support for growing reuse in our analysis. First, we find that the most popular data and methods links in papers are heavily reused across papers (Fig. 12). For instance, the most popular 1% of methods links in physics, computer science, and math account for 31.1%, 16.5%, and 7.8% (respectively by field) of the total mentions of methods links among papers in our dataset. Similarly, the most popular 1% of data links in physics, computer science, and math account respectively for 37.3%, 18.3% and 9.6% of the total data link mentions. The most popularly referenced data and method links are shown in Appendix Table 2 and 3.

Second, we find the rate at which papers reuse previously introduced data and methods links is also growing (Fig 3(a) and Fig 3(b)). For instance, computer science papers on average reused 0.02 data links in 2011 and up to 0.10 data links in 2020. Similarly, computer science papers mention 0.04 methods links in 2011 and up to 0.16 reused methods links in 2020. Comparable growth rates of reused links can be observed in physics as well, while the reuse rate in math is relatively stable.

[∥] https://github.com/allenai/s2orc-doc2json

[**] In their taxonomy, links to data is referred as 'material' link.

[††] Note that the same link can be be classified into different categories depending on its usage context, e.g., one paper may mention that they use the data from a URL (thus the URL will be classified as a data link for this context) while another paper may mention that they use the code from the same URL (thus this URL will be classified as a methods link for this second context)

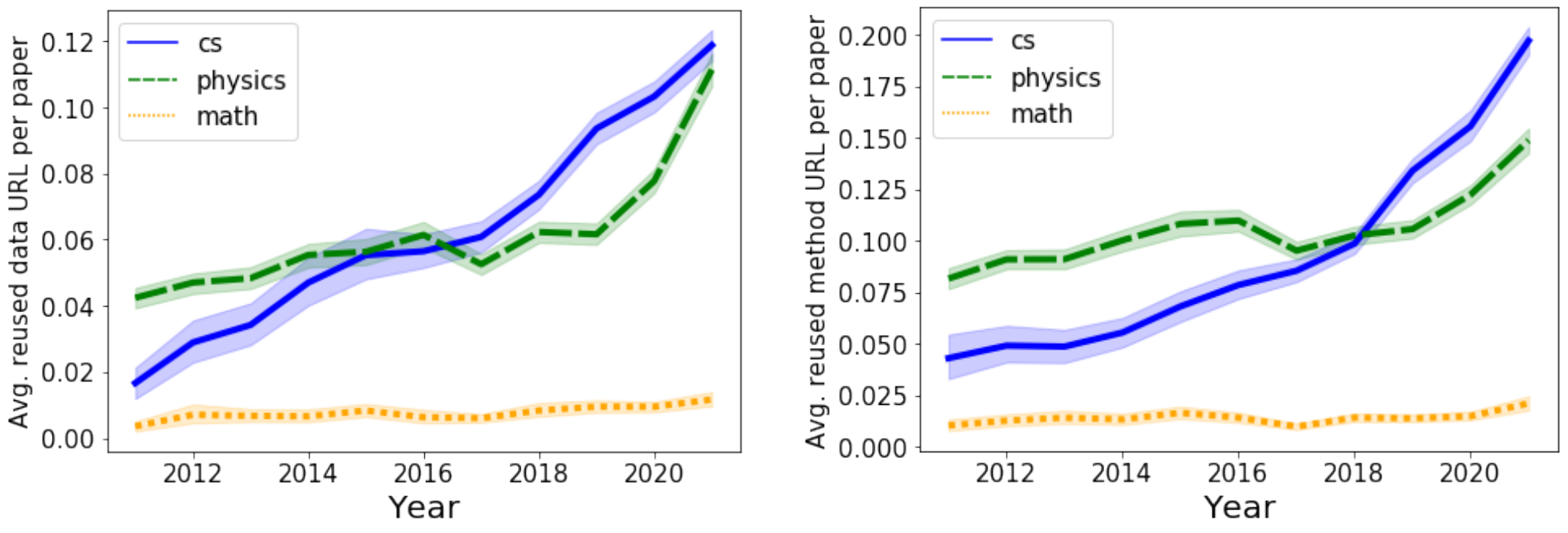

(a) Average number of reused data URLs per paper over time.

(b) Average number of reused method URLs per paper over time.

**Fig. 3.** Data and methods links in papers are increasingly reused over time in different fields, especially computer science. Reusing data and method links is most common in physics.

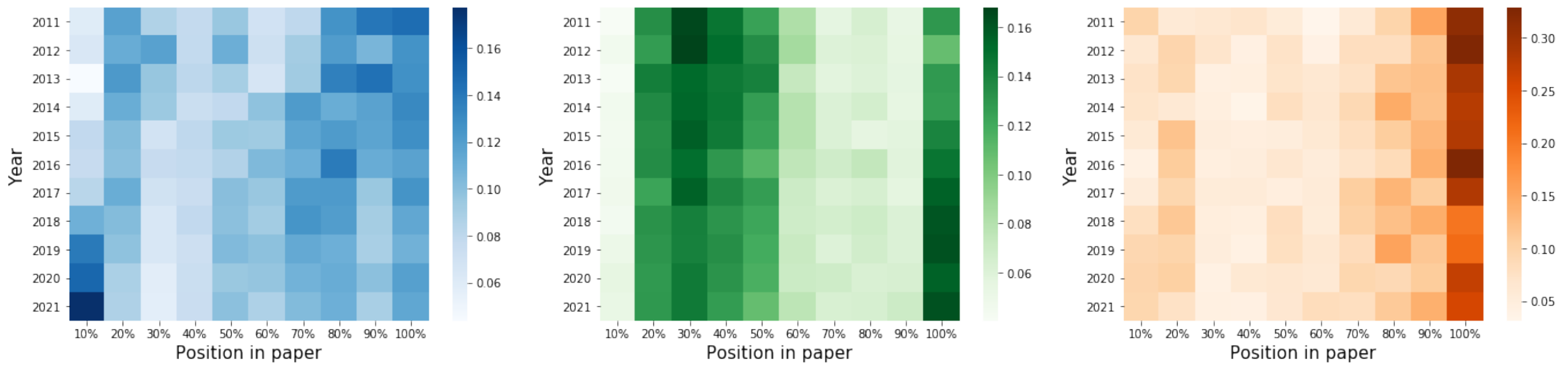

(a) CS

(b) Physics

(c) Math

**Fig. 4.** Different fields standardize where researchers put links in their articles. The figures here show the likelihood of methods links appearing in different parts of papers (from first 10% to last 100% of text) by year for each of our different fields. Darker colors indicate higher probabilities of finding methods link in corresponding positions for a given year. We find that CS papers initially place methods links in the last 100% of text and increasingly shift toward listing methods links in the first 10% of their papers. Physics flips in the opposite manner, shifting from start to ending placement over time. And math is most likely to place methods links in last 100% of the paper.

These results suggest that computer science and physics scholars are increasingly replicating or building upon data and methods resources presented in earlier works.

Among the three fields, physics has a significantly higher proportion of data and methods link reuse compared to the other fields. As shown in Figures 13(a) and 13(b)), close to 60% of total data and methods link mentions in physics reuse data and methods links introduced in prior work. This is a much higher than observed in math and computer science. The difference may reflect physics' distinctive reliance on shared datasets and material resources, such as joint astronomy databases like SDSS.

**Data and methods link sharing practices have become more standardized and centralized.** We observe that data and methods link sharing practices are increasingly standardized, indicating that the involved research communities may have formed expectations or norms around where and how researchers should share data and methods in publications.

First, there is increasing consensus on where links should be presented in papers. This increased consensus indicates set norms and procedures which help open science practices spread more easily. Specifically, we conduct analysis on the positions within paper where people hold data and method links, and find that different fields have emerged their own 'standardized' places of holding links, e.g. math and physics tend to hold links towards the end of the paper, while computer science tends to present them in the very beginning (Fig 4 and Appendix D). The results indicate that method and data link sharing is not only becoming more common practice, but also formed standard usage patterns.

Second, we find the domains hosting methods and data links have become increasingly standardized over time. As shown in Fig 2, we calculate the Gini coefficient—a metric that measures the spread/concentration of a distribution—of associated domains in data and methods links each year by field. We find that the Gini coefficient is increasing over time for all three fields, indicating growing concentration of links hosted in fewer domains. Through further investigation (Appendix B), we find that certain institutional domains (e.g. harvard.edu, nasa.gov) have been consistently popular over time while domains such as github.com and huggingface.co have increased in popularity in recent years. For instance, in computer science, 12 methods URLs are hosted on github.com in 2011, growing to 1,732 in 2016, and 19,718 in 2021, making it the most popular domain associated with methods sharing in computer science over the span of our analysis. The use of huggingface.co as a domain to host methods has also grown exponentially in recent years; the first appearance of this domain was in 2019, when it was

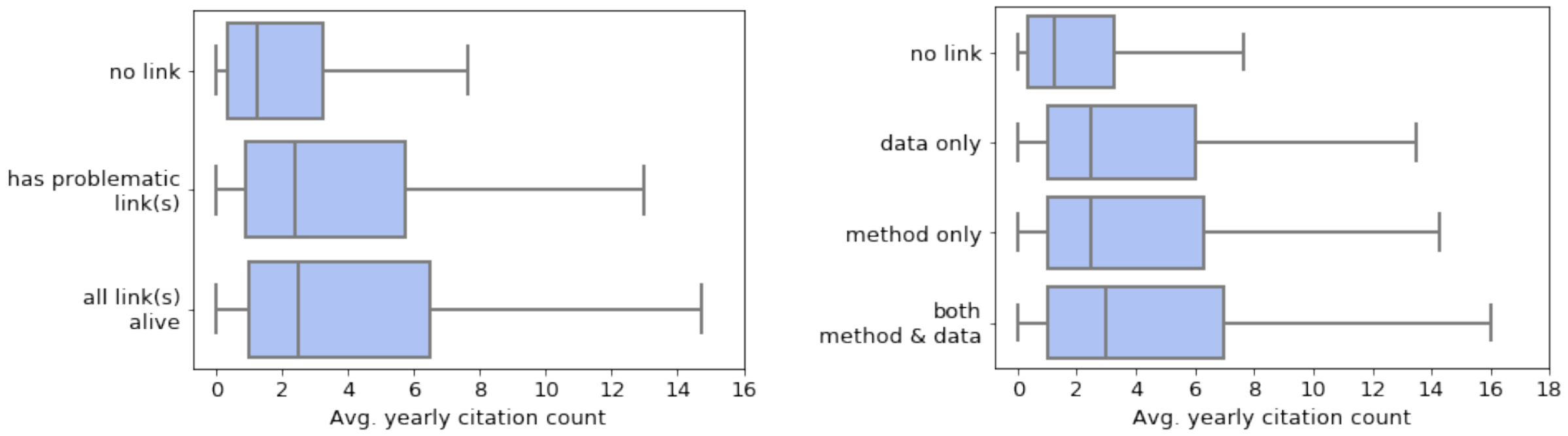

(a) Average annual citation count for papers with different types of links.

(b) Average annual citation count for papers with different retrievability status of links.

**Fig. 5.** Sharing data and methods links and sharing links that are alive are associated with higher paper citation counts. These box plots show the quartiles of average annual citation counts received by papers with different link types and retrievability status.

mentioned 13 times, and by 2021, it was mentioned 673 times.[‡‡]

**Standardized open science practices avoid link rot and sustain greater data and methods access over time.** Scholars have increasingly centralized their data and methods links in certain web domains. One possible reason they are drawn to these domains is because they maintain active links for longer and reduce the likelihood of link rot. Regression analyses (Details in Appendix E) support this. Our analysis indicates that data and methods links appearing in popular domains are more likely to be alive than problematic over time, indicating the benefits of standardizing link hosting to select well-resourced domains. Our analysis also shows that if the popularity (count) of a domain that hosts links doubles, the odds of a link remaining alive will be 108% the original; if the popularity (count) of a specific URL doubles, the odds of the link remaining alive will be 118% the original.

**Sharing data and methods links, and live versions of them, is associated with higher paper citation count.** We further investigate whether the act of link sharing is associated with downstream measures of impact, where we focus on the number of citations a paper receives.

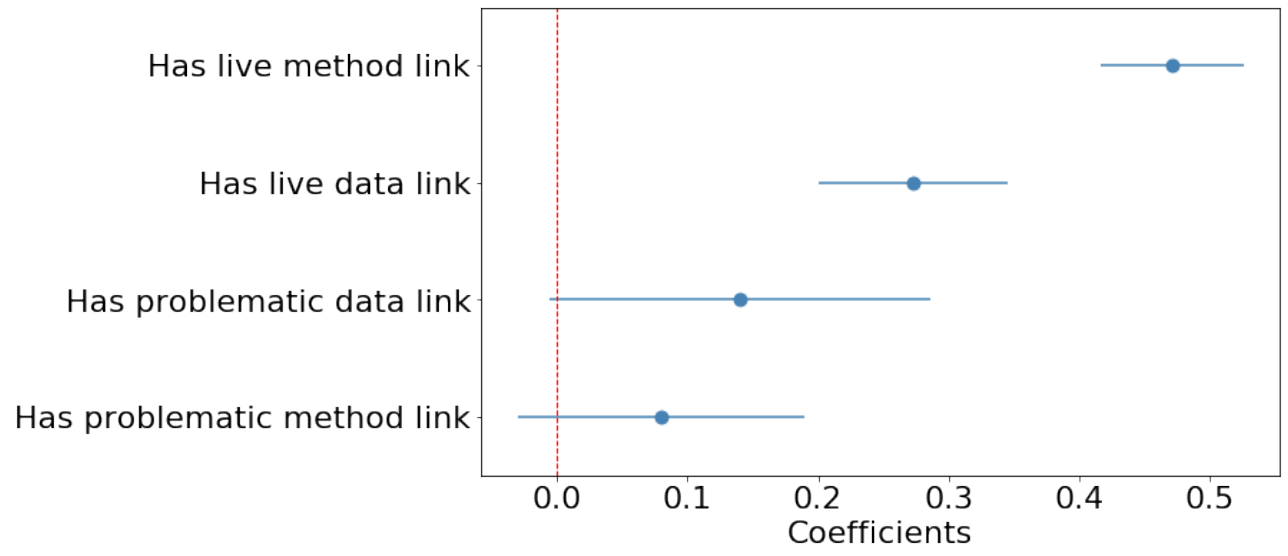

**Fig. 6.** Coefficients of negative binomial model on predicting citation count from paper link type and status. Sharing data and methods links, especially those that are alive and retrievable, is associated with higher paper citation counts.

Fig 5(a) compares the number of average annual citation counts for papers with no data or methods links, with data link(s) only, with methods link(s) only, and those with both methods and data links. As observed, papers with links to data and methods receive higher annual citations than papers with no data or methods links. Fig 6 compares the number of average annual citations received by papers with no data or methods links, papers with links that are no longer retrievable, and papers with all links alive and accessible. We observe that papers with data and methods links, whether or not these links are alive or problematic, receive more annual citations than papers with no data or methods links.

Next, we formally test whether certain types of links correspond to heightened citation count, net of other factors. We ran a negative binomial regression model with paper citation count as the dependent variable. We use four key independent variables of interest: whether the paper has a live methods link; live data link; problematic data link; or problematic methods link, controlled by the paper's field of study and the age of the paper (details in Appendix F). As shown in Fig 6, we find that link provision is valued and rewarded in these scientific communities. Having a live methods link in a paper corresponds to an increased citation count of 60.0% ($p<0.001$); a live data link with an increased citation count of 31.0% ($p<0.001$), in comparison to papers with no links; and having problematic methods and data links in a paper are associated with increased citation counts of 8.3% and 15.0% in comparison to papers with no links, though the relationship is not statistically significant. While some data and methods URLs are prone to link rot, both live links and problematic links are associated with higher citation count, and papers with links that are alive at the time of analysis are associated with the highest citation counts. In other words, the act of link sharing is associated with higher impact compared to no link sharing overall, but having a link at which to retrieve actual data or code implementations is associated with even higher citation counts. Our results are in line with what has been observed in prior work analyzing papers from select publication venues or papers with data availability statement (18, 19).

[‡‡]This annual number continues to grow; in the first half of 2022 through June, huggingface.co is mentioned 686 times.

## Discussion

Our work provides large-scale evidence that the practice of openly sharing and reusing data and methods in scientific preprints is on the rise. Data and methods link sharing in science is increasingly standardized and centralized in certain URL domains which help to sustain active links. The increasing practice of open science through link sharing is also associated with higher research impact. While prior work has reported preliminary empirical evidence on the citation advantages of open data practices through analysis of data availability statements (19) or carefully curated data sets (6, 16), they focus on specific (prestigious) venues or research topics and seldom consider methods sharing separately. Thus, it is unclear to what extent their conclusions generalize to the broader scientific community. Our work provides an automated and scalable approach that complements prior efforts at extracting data and methods sharing mentions in text by automatically extracting and classifying URL mentions in the full text of papers. Results from our study offer further evidence on the increased diffusion and downstream benefits of data and methods sharing in academic publishing. Our conclusions suggest that scientific communities are increasingly practicing authentic forms of open science via link sharing, not only providing access to sources of data and methods, but reusing them in ways suggesting increased replication. Among the fields we study, authors are also adopting new practices to facilitate these efforts, such as increased usage of standardized web domains for link hosting and rewarding scientific work that adopts such practices with increased citation count.

We find that data and methods links hosted on popular domains are more likely to persist and remain retrievable, thus suggesting that the scientific community may want to encourage link sharing through more standardized and well-resourced domains. For example, publishers may want to provide guidelines to researchers encouraging usage of centralized data repositories such as Zenodo or Figshare, or commercial solutions such as GitHub. These domains are more likely to be maintained, compared to less popular sites such as personal websites, where linked data and methods resources may be lost over time. Our findings also suggest the need to maintain long-term availability of shared data and methods links, as ours and others' results show that a large proportion (over 15% data and method links from 2011) of links referenced in papers fall to link rot (30).

From a tool building perspective, our work suggests the possibility that automated methods could help preserve important paper artifacts. Link rot hinders reproducibility and diminishes the reuse of scientific resources. Publishers and preprint servers could take advantage of our pipeline during the review process of papers to automatically archive data and methods artifacts. For example, when a researcher publishes their paper, one could identify URL links, automatically retrieve the content of these links, and deposit them in public repositories for future reference. Under such a process, fewer important scientific artifacts would be lost to link rot.

In sum, our work develops an NLP-based automated pipeline to analyze the adoption of data and methods sharing practices in scientific papers by extracting and identifying data and methods links through the context of link mentions. This opens up the possibility to evaluate forms of open science practices in a scalable way. Our work is limited in several ways: 1) the link classification algorithm currently supports only a single class classification for a specific link mention context.[§§] While most links only involve one type of artifact, i.e., data, methods, or supplement, there are scenarios where both data and methods are included in the same link under the same context (e.g., 'we release our code and data at [URL]'). Our current model only classifies these mention contexts into a single class.[¶¶] Nevertheless, our approach makes accurate distinctions between data and methods links as opposed to supplement links, which are less central and important to the practice of open science. Based on the infrequency of multi-class contexts, we believe that implementing multi-class classification would not alter our main findings and thus leave this to future work. 2) Our current analysis is also limited to the fields of computer science, physics, and mathematics. These are the most well represented fields in arXiv, which provides the source of documents from which we can more accurately identify and extract URLs (due to access to LaTeX source). Future work can extend our methods to PDF documents (33, 36), which would enable analysis on papers from other publication venues and preprint servers. However, we note that parsing and extracting URLs from PDF documents poses additional challenges and is likely to be less accurate. 3) Moreover, our regression analysis indicates correlation, not causality. We believe the description of relationships and trends is an important first step in understanding whether and how open science is practiced, and further work is needed to study if open science interventions can cause downstream differences in paper impact. 4) Finally, a link being live is not equivalent to the data or methods being accessible at that web URL. Our current work does not evaluate the actual content at the webpage on the other end of the link, and thus places an upper bound on the estimate of live links. Integrating our pipeline with automated methods that can discern the content on the resulting webpage, or human evaluation to determine whether data and methods are actually accessible, would be a fruitful direction for future work. We hope our methods and analyses will inspire other work on this topic as well as the promotion of open science and reproducibility practices in computational fields and beyond.

[§§] Note that our approach can differentiate the same link mentioned for different purposes in multiple contexts (e.g., we are able to differentiate a link used as a data link in one paper but as a method link in another paper), as discussed in the Methods section.

[¶¶] Such same-context multi-class scenarios are rare—only 2 out of 100 link mention records in our validation set (the 100 link mention contexts randomly sampled from our analyzed dataset that we carefully labeled to validate our link classification method) should be labeled as multi-class for the specific context (in both cases the link should be labeled as both a data and a methods link); the other 98 link contexts referenced only a single link type.

**ACKNOWLEDGMENTS.** Hancheng Cao is supported by the Stanford Interdisciplinary Graduate Fellowship. This work was supported in part by NSF Grant 2033558. We thank Bryan Newbold for answering questions about HTTP requests and web archiving. We thank the Semantic Scholar and AllenNLP research teams for feedback on the project and manuscript.

## Code Availability

The codes can be accessed at https://github.com/caohanch/paper_data_method_sharing/.

**Table 2. Top data URLs in computer science, physics and math, and the proportion of mentions they account for in each field.**

| Computer Science | Physics | Math |
|---|---|---|
| yann.lecun.com/exdb/mnist (0.50%) | cosmos.esa.int/gaia (3.4%) | csie.ntu.edu.tw/ cjlin/libsvmtools/datasets (1.1%) |
| grouplens.org/datasets/movielens (0.32%) | cosmos.esa.int/web/gaia/dpac/consortium (3.2%) | cosmos.esa.int/gaia (0.85%) |
| nlp.stanford.edu/projects/glove (0.30%) | ned.ipac.caltech.edu (1.5%) | cosmos.esa.int/web/gaia/dpac/consortium (0.81%) |
| jmcauley.ucsd.edu/data/amazon (0.30%) | leda.univ-lyon1.fr (1.2%) | yann.lecun.com/exdb/mnist (0.80%) |
| code.google.com/archive/p/word2vec (0.24%) | simbad.u-strasbg.fr/simbad (0.60%) | csie.ntu.edu.tw/ cjlin/libsvmtools/datasets/binary.html (0.46%) |

**Table 3. Top method URLs in computer science, physics and math, and the proportion of mentions they account for in each field**

| Computer Science | Physics | Math |
|---|---|---|
| pytorch.org (0.46%) | iram.fr/IRAMFR/GILDAS (0.92%) | math.uiuc.edu/Macaulay2 (0.24%) |
| github.com/huggingface/transformers (0.34%) | dirac.ac.uk (0.59%) | sagemath.org (0.20%) |
| spacy.io (0.31%) | astropy.org (0.5%) | github.com/deel-ai/deel-lip (0.17%) |
| tensorflow.org (0.28%) | casa.nrao.edu (0.43%) | dirac.ac.uk (0.16%) |
| github.com/google-research/bert (0.28%) | iraf.noao.edu (0.42%) | cvxr.com/cvx (0.15%) |

## A. Details on URL extraction and normalization

We adapt a regular expression from Stack Overflow for use in URL detection. The complete expression we use is:

```
((?:https?://)?(?:(?:www\.)?(?:[\da-z\.-]+)\.(?:[a-z]
{2,6})|(?:(?:25[0-5]|2[0-4][0-9]|[01]?[0-9][0-9]?)\.)
{3}(?:25[0-5]|2[0-4][0-9]|[01]?[0-9][0-9]?)|(?:(?:[0-
9a-fA-F]{1,4}:){7,7}[0-9a-fA-F]{1,4}|(?:[0-9a-fA-F]{1
,4}:){1,7}:|(?:[0-9a-fA-F]{1,4}:){1,6}:[0-9a-fA-F]{1,
4}|(?:[0-9a-fA-F]{1,4}:){1,5}(?::[0-9a-fA-F]{1,4}){1,
2}|(?:[0-9a-fA-F]{1,4}:){1,4}(?::[0-9a-fA-F]{1,4}){1,
3}|(?:[0-9a-fA-F]{1,4}:){1,3}(?::[0-9a-fA-F]{1,4}){1,
4}|(?:[0-9a-fA-F]{1,4}:){1,2}(?::[0-9a-fA-F]{1,4}){1,
5}|[0-9a-fA-F]{1,4}:(?:(?::[0-9a-fA-F]{1,4}){1,6})|:(
?:(?::[0-9a-fA-F]{1,4}){1,7}|:)|fe80:(?::[0-9a-fA-F]{
0,4}){0,4}\%[0-9a-zA-Z]{1,}|::(?:ffff(?::0{1,4}){0,1}
:){0,1}(?:(?:25[0-5]|(?:2[0-4]|1{0,1}[0-9]){0,1}[0-9]
)\.){3,3}(?:25[0-5]|(?:2[0-4]|1{0,1}[0-9]){0,1}[0-9])
|(?:[0-9a-fA-F]{1,4}:){1,4}:(?:(?:25[0-5]|(?:2[0-4]|1
{0,1}[0-9]){0,1}[0-9])\.){3,3}(?:25[0-5]|(?:2[0-4]|1{
0,1}[0-9]){0,1}[0-9])))(?::[0-9]{1,4}|[1-5][0-9]{4}|6
[0-4][0-9]{3}|65[0-4][0-9]{2}|655[0-2][0-9]|6553[0-5]
)?(?:/[\w\.-]*)*/?))
```

As such, links that have URL like patterns will be extracted. The regular expression ensures that special characters from the main text will not be included as part of the extracted result. e.g., a URL followed by a comma within the sentence.

While most of the extracted URLs are of good quality, the extraction left a number of issues: 1) sometimes only part of the URL is extracted as there may be substrings that contain special characters, which is quite common in links from YouTube and Google Drive. For instance, instead of https://www.youtube.com/watch?v=G9llFqAwI-8 only https://www.youtube.com/watch was extracted, and 2) there can be URL variations pointing to the same site, e.g., a URL including an http header and one that does not.

To resolve the first issue, for each of the extracted URL from our initial extraction, we correct the extracted URL by checking and appending any additional 'suffix' substring that consists of English letters and special characters (excluding white-space characters) after the initial extraction using the regular expression. In total, 62,040 records are corrected by appending substrings that were not included in the initial extractions. To resolve the second issue, we further process all extracted and corrected URLs using a URL normalization package*** and remove the prefixes `http://`, `https://` and `www`.

Finally, we manually validate our link extraction approach. 1) We randomly sample 100 records from our dataset and manually review whether the final extracted URLs after normalization match the URLs found in the original paper. We find that 97 our of 100 records are correct, and the other 3 include spaces in the original URLs that were difficult to automatically correct. This validates the overall high quality of our extraction method. 2) We also randomly check 100 records that have different initial extracted URLs versus corrected URLs, and find that 99 of the URLs identified as having initial extraction issues are indeed corrected by our followup method, demonstrating the quality and necessity of correction.

## B. Top URLs, domains, and suffixes

We explore where researchers host their artifacts. Fig 7 and Fig 8 show the top 10 domains and URLs researchers mention in aggregate in papers in our dataset, split by link type and field of study.

The most popular domain is github, which is used frequently across different fields but especially popular for computer science. In comparison, domains such as nasa, esa, caltech, harvard, stsci, and sdss are primarily used in physics. Interestingly, most top unique URLs mentioned in arXiv papers are in physics, indicating the high reusability of certain databases and resources in that field. We further present the most popular URLs and domains broken down by fields in Tab 2, Tab 3, Fig 9, and Fig 10.

*** https://github.com/iipc/urlcanon

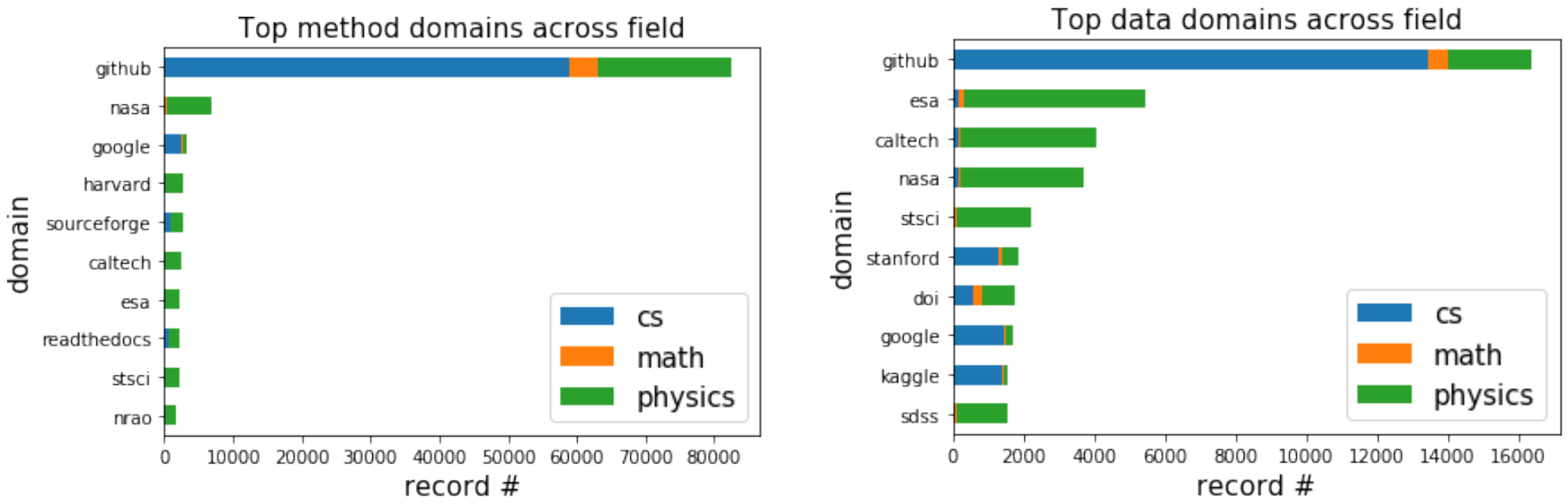

(a) Top domains across fields for method links.

(b) Top domains across fields for data links.

**Fig. 7.** Top domains for method and data links.

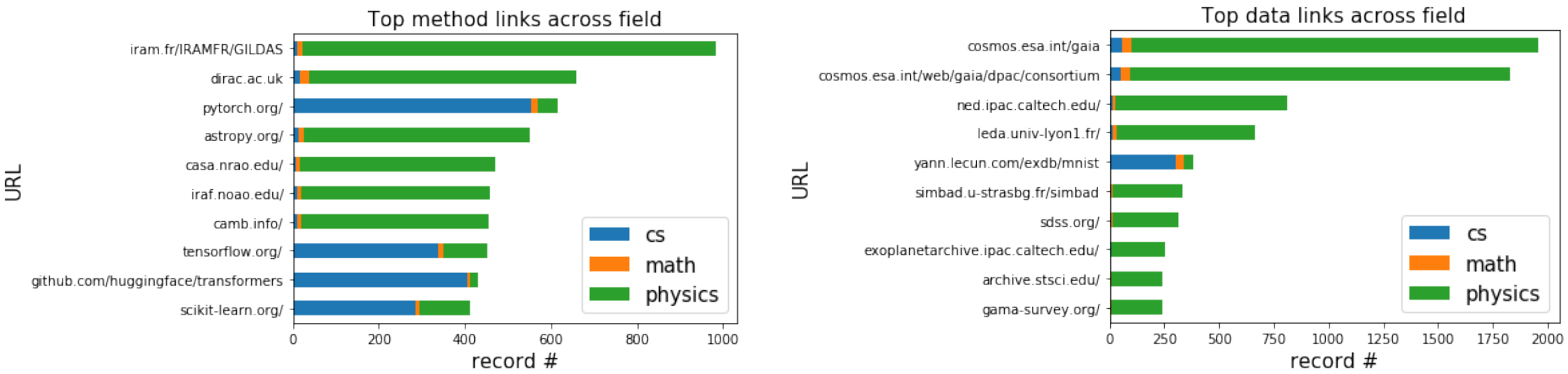

(a) Top method URLs.

(b) Top data URLs.

**Fig. 8.** Top method and data URLs.

## C. Supplementary results on link usage and re-usage over time

Fig 11 shows data and methods link usage over time, as a proportion of all links. Computer science and physics have growing proportions of data and methods link sharing over time. Mathematics, on the other hand, persists in mostly using supplement type URLs. Fig 12 shows the distribution of percentiles of data and methods links that are reused in our dataset. As can be seen, most data and methods links shared in all three fields are mentioned in other papers; e.g., the 60th percentile of computer science data-type links make up close to 100% of data link mentions.

We further provide statistics of overall usage and re-usage of all links (contains all three types: data, methods, and supplement) over time in, as a supplementary to the main results on data and method link sharing. Similar to what is observed over data and method links, there have been increasing usage (Fig 14) and reusage (Fig 16) of links in all three fields over time, with especially notable increase in computer science. Popular links constitute most of the mentions in the three fields (Fig 17). Moreover, the domains of links are becoming more concentrated over time across fields (Fig 15).

## D. Where are data and methods links located in papers?

We explore whether there are norms around specific locations in papers where authors are more likely to link to data and methods artifacts, e.g., abstract/introduction, data availability statement, appendices/supplementary materials. To enable such analysis, for each paper with links, we partition them into 10 equal-sized bins based on paragraph position in the paper, and indicate whether a data or methods link lies within each bin (e.g., if a data link is in the abstract, it is typically binned into the first 10% of the paper). These aggregated trends are shown in Fig 18 (data links) and Fig 4 (methods links).

Mathematics and physics tend to have data and methods links in the final 10% of their papers (especially physics for data and mathematics for methods). Manual inspection shows these are likely to be found under section headings such as ‘data availability,’ ‘method availability,’ ‘conclusion,’ or ‘appendix.’ Links in computer science papers, on the other hand, tend to concentrate in the very beginning (‘abstract’) of papers (for methods) and more towards the middle/end of papers (for data).

Some of these emergent archetypal data and methods sharing practices have become more the norm in recent years. For instance, 6.0% of computer science method links appearin the first 10% of the paper in 2011, but more than 17.8% of method links in computer science follow this archetype after 2021. Similarly, in physics, presenting data links in the final 10% of the paper has emerged as a norm, with more than 36.2% of data links mentioned at end of the paper in 2021, compared to 15.6% in 2011.

## E. Are data and methods links retrievable?

We analyze the extent to which links in papers remain retrievable over time. Web URLs may become inaccessible after some time, a phenomenon commonly known as “link rot.” Prior work has shown that dead links are common in scientific articles (30). As demonstrated by Fig 19, links across fields show consistent patterns of being more accessible in recent years while older links are more likely to be problematic (subject to link rot and other issues such as an artifact moving to a new location on the web). Around 15% of links from 2011 are problematic while less than 5% of links are problematic in 2021. We identify problematic links by studying the HTTP status code that is returned when a link is requested. We use the python requests library to make URL requests, which returns HTTP status codes as well as error codes associated with unreachable URLs. In Tab 4, we summarize the distribution of link retrieval status (HTTP codes and request errors) for data and methods links in our dataset.

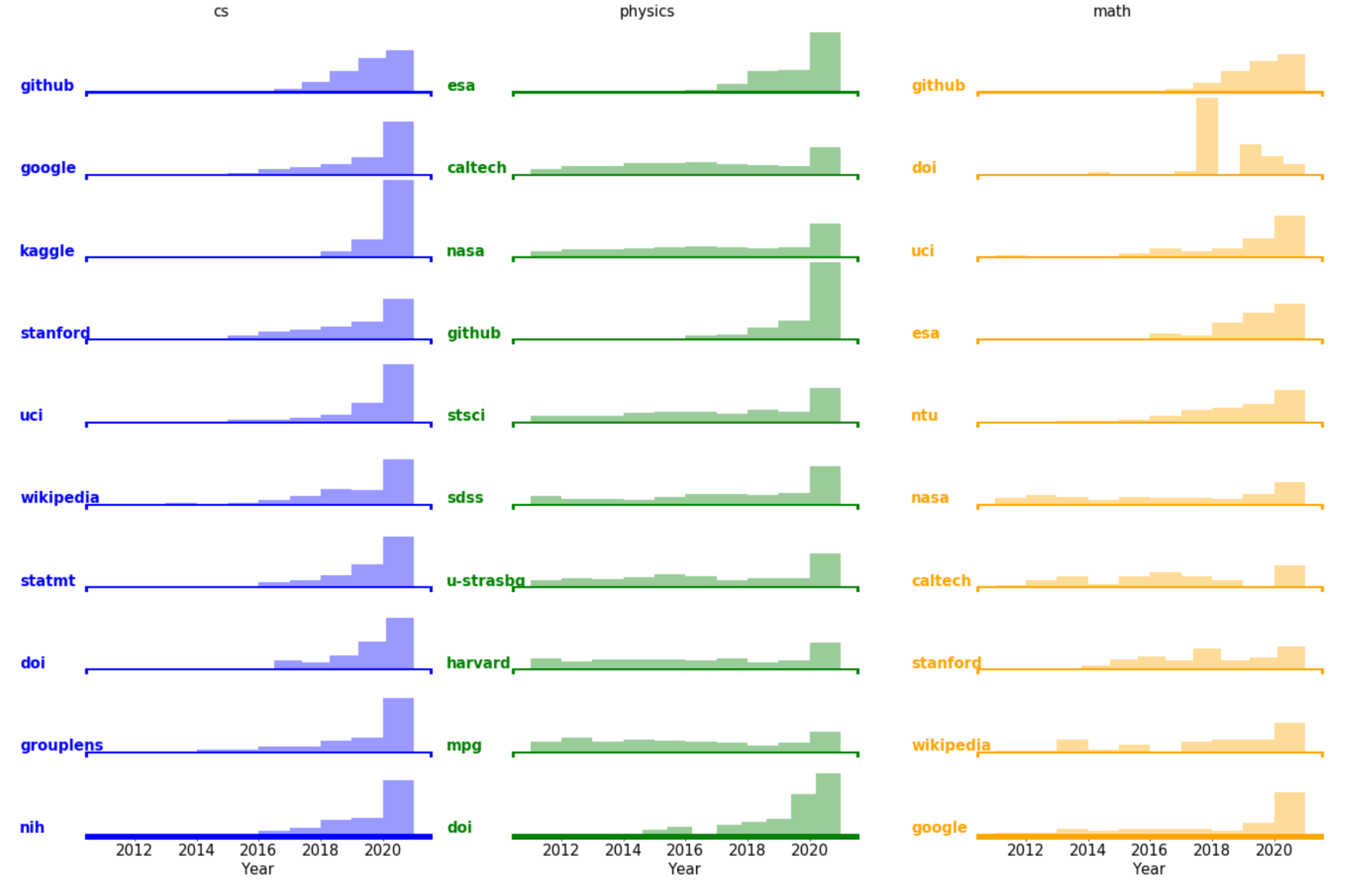

**Fig. 9.** Top domains that hold data links in CS, physics and math over time.

## F. Details of regression

We investigate what features of data and methods links are associated with retrievability. We run a logit regression (coefficients in Tab 5), with a binary variable indicating whether a link is live (200 status code) or problematic (non-200 status code) as the dependent variable; and field, popularity of the URL (log-transform of URL count), popularity of the web domain (log-transform of web domain count), the paper age, the square of the paper age, the type of the artifact, and the citation count of the paper (log-transformed) as independent variables. We find that popular links, links used in highly cited papers, links hosted at popular domains, and links referenced in recent years are more likely to be alive than others. Data and methods links are also more likely than supplement links to be alive. Specifically, if the popularity of a web domain doubles, the odds of returning a live link will be 108% the original; if the popularity of a URL doubles, the odds of getting a live link will be 118% the original.

We ran a negative binomial regression model with paper citation count as the dependent variable, with four key variables of interest: whether the paper has a live data link, whether the paper has a live methods link, whether the paper has a problematic data link, and whether the paper has a problematic methods link. We control for field of study, the paper age, and the square of the paper age. The regression result is shown in Table 6. The regression shows that having a live methods link in a paper corresponds to an increased citation count of 60.0% ($p$<0.001); a live data link with an increased citation count of 31.0% ($p$<0.001), in comparison to papers with no links; and having problematic methods and data links in a paper are associated with increased citation counts of 8.3% and 15.0% in comparison to papers with no links, though the relationship is not statistically significant.

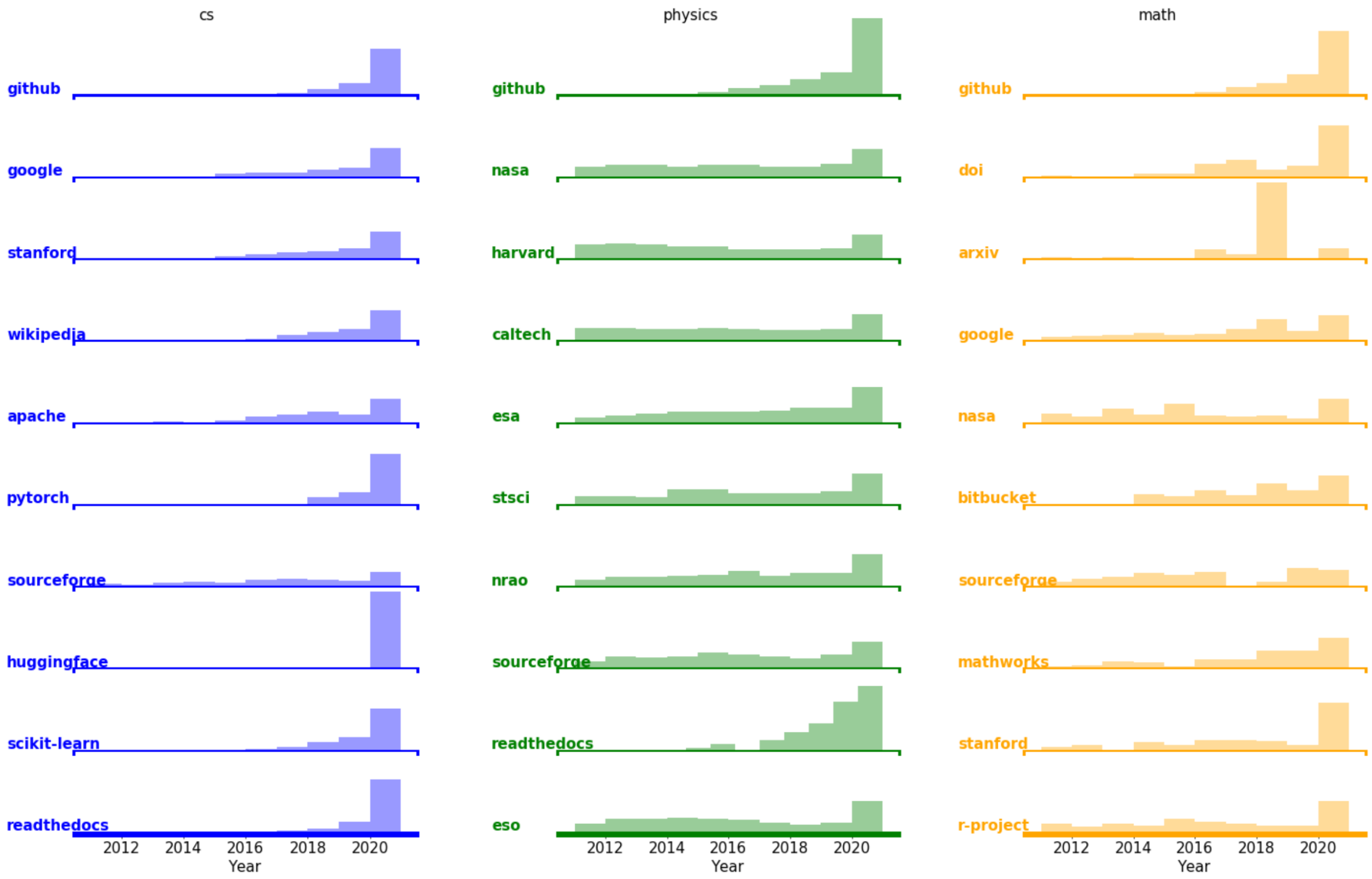

**Fig. 10.** Top domains that hold method links in CS, physics and math over time.

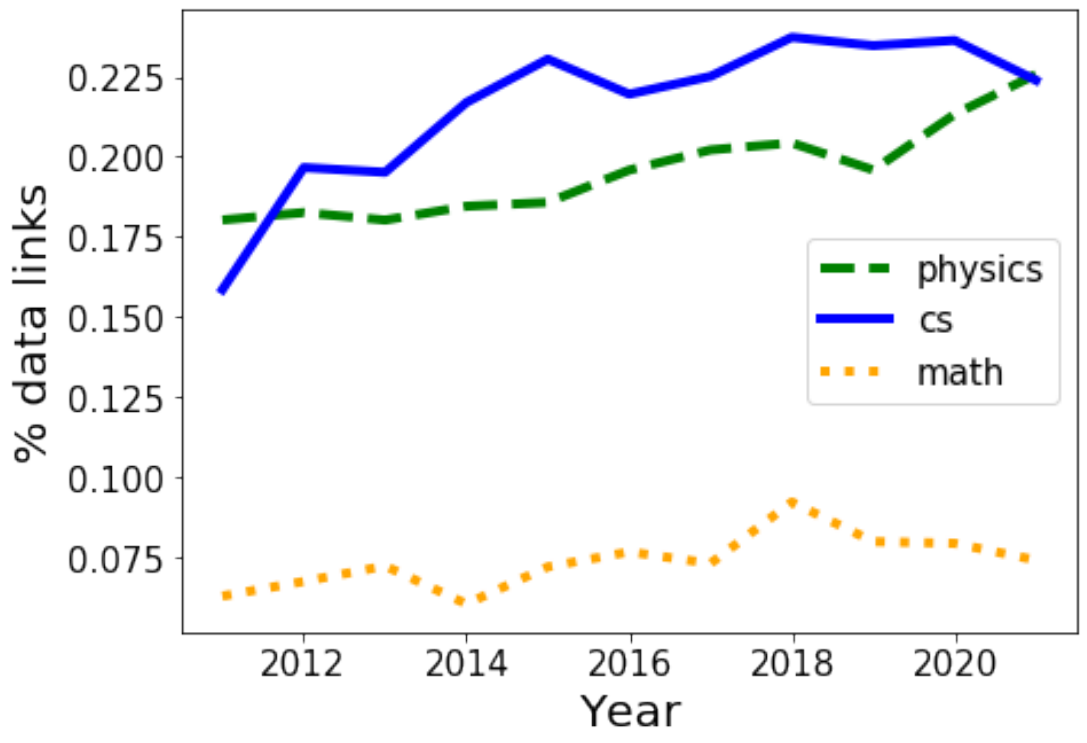

(a) Proportion of URLs belonging to data type over time.

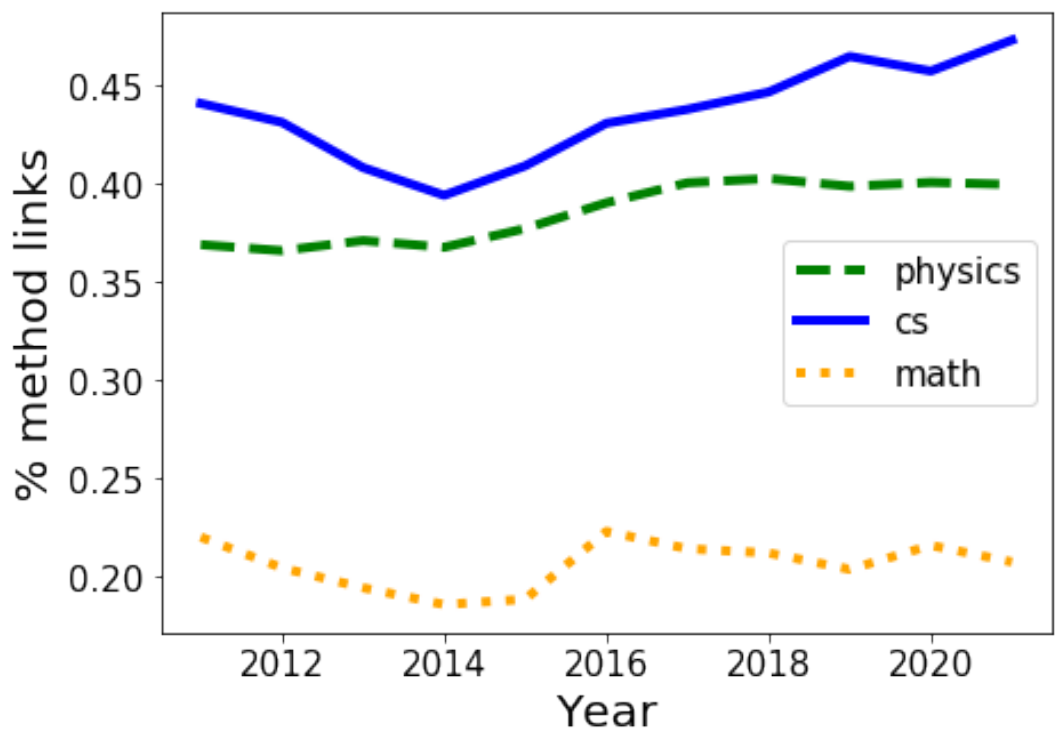

(b) Proportion of URLs belonging to method type over time.

**Fig. 11.** Computer science and physics have high proportion of method sharing links, and increasingly provide data links. Math on the other hand persists in mostly using supplemental URLS.

**Table 4. URL links by retrievability status**

| HTTP status code / requests error | Description | Number of links (proportion) |
|---|---|---|
| 200 | Success | 241,082 (83.1%) |
| 404 | Not Found | 22,208 (7.7%) |
| ConnectionError | A Connection error occurred | 8,548 (2.9%) |
| 403 | Forbidden | 4,657 (1.6%) |
| SSLError | A website cannot provide a secure connection, e.g., not having an SSL certificate | 4,265 (1.5%) |
| ConnectTimeout | The request timed out while trying to connect to the remote server | 3,157 (1.1%) |
| 503 | Service unavailable | 2,441 (0.84%) |
| 429 | Too Many Requests | 1,686 (0.58%) |
| ReadTimeout | The server did not send any data in the allotted amount of time | 943 (0.33%) |
| 500 | Internal server error | 403 (0.14%) |

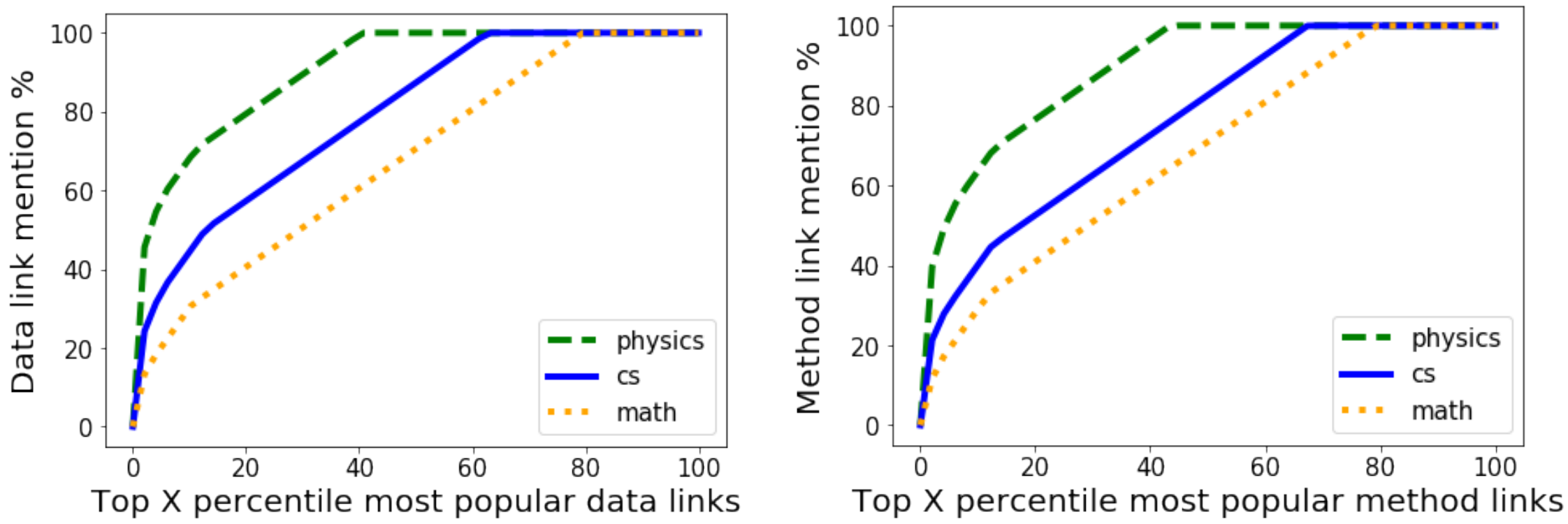

(a) Popular data links are heavily reused

(b) Popular method links are heavily reused

**Fig. 12.** Distribution of data and method link reusage.

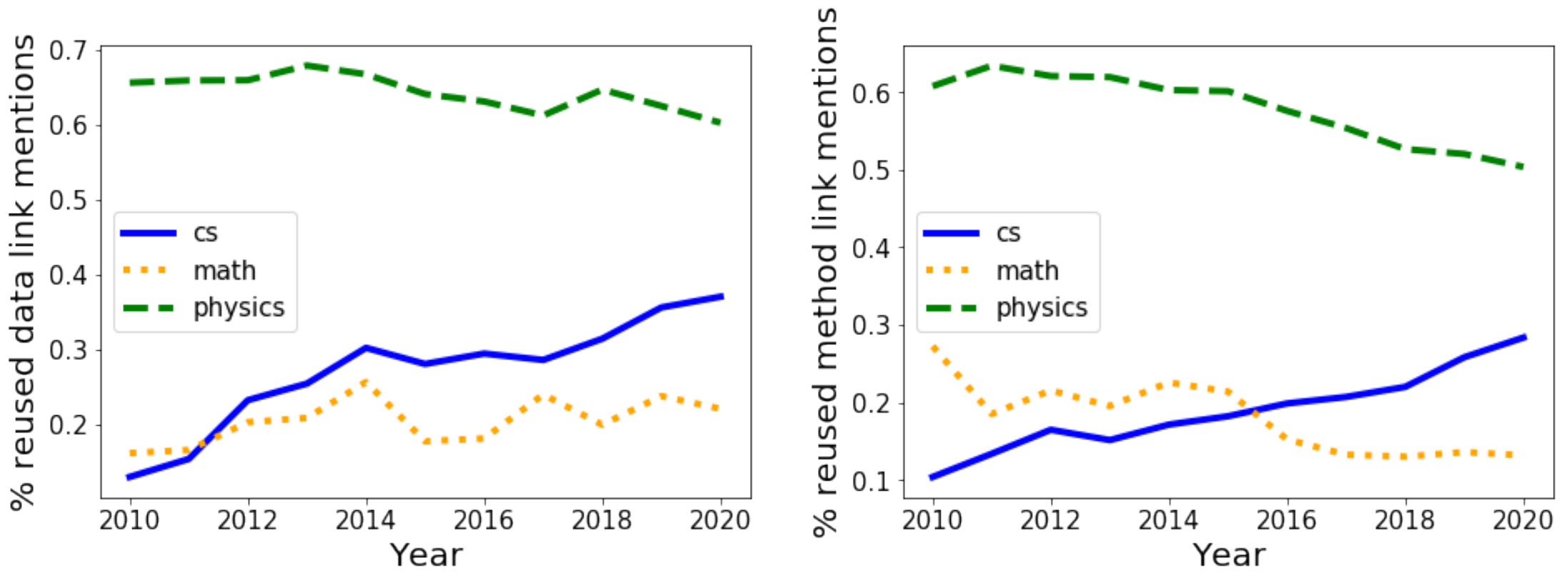

(a) Proportion of reused data links among all data link usage over time.

(b) Proportion of reused method links among all method link usage over time.

**Fig. 13.** Proportion of reused data and method links over time.

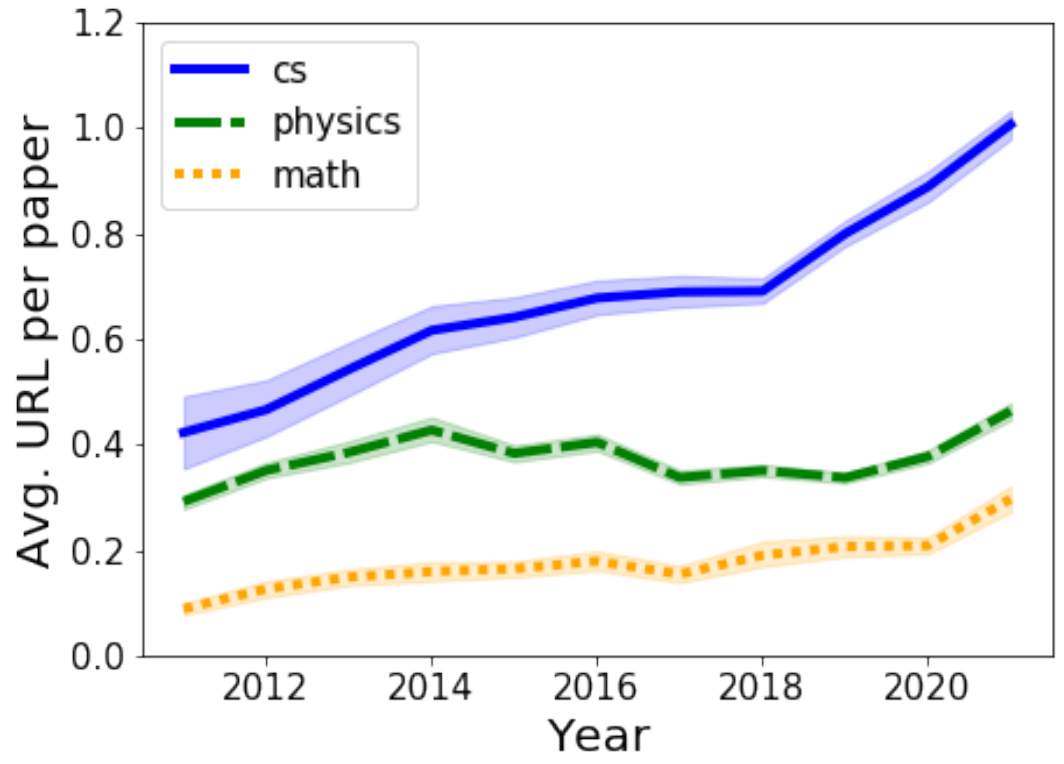

**Fig. 14.** Increasing number of links have been provided in papers over years across fields.

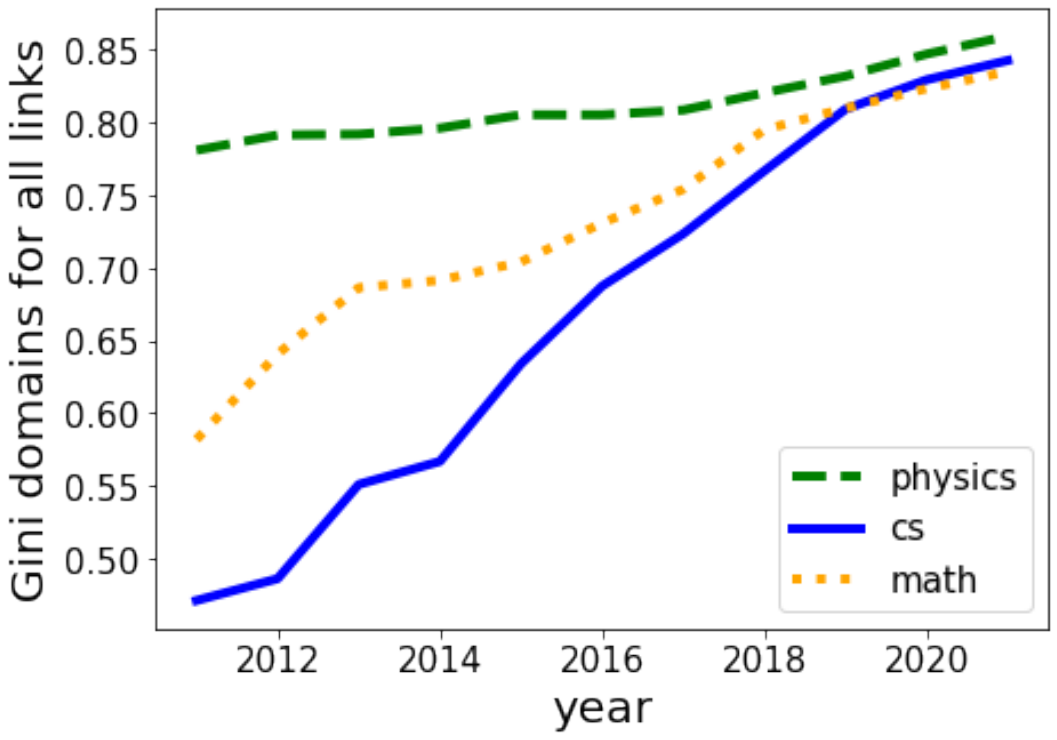

**Fig. 15.** Domains across all types of links become more concentrated over time across field.

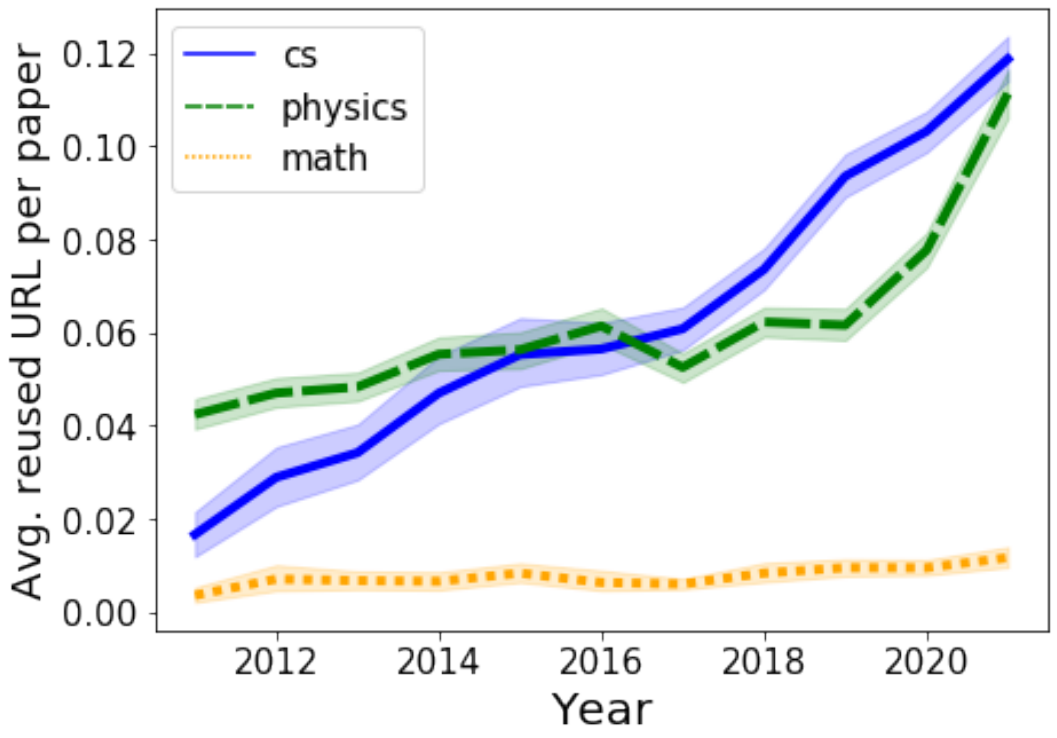

**Fig. 16.** Number of reused links per paper over time.

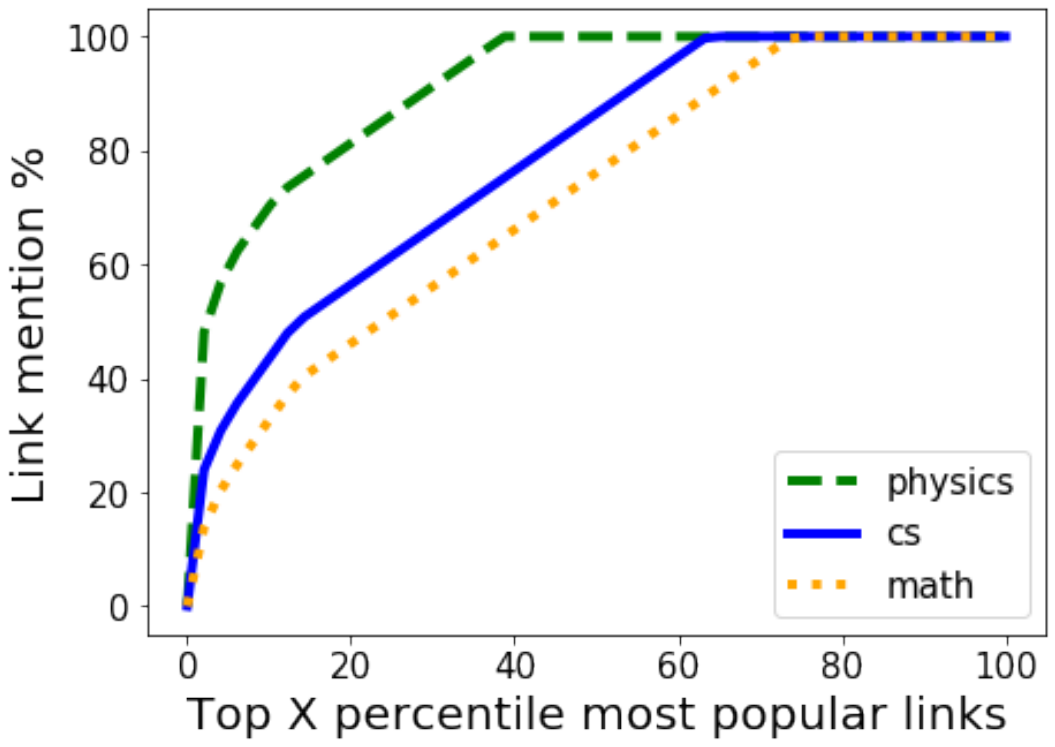

**Fig. 17.** Popular links are heavily reused.

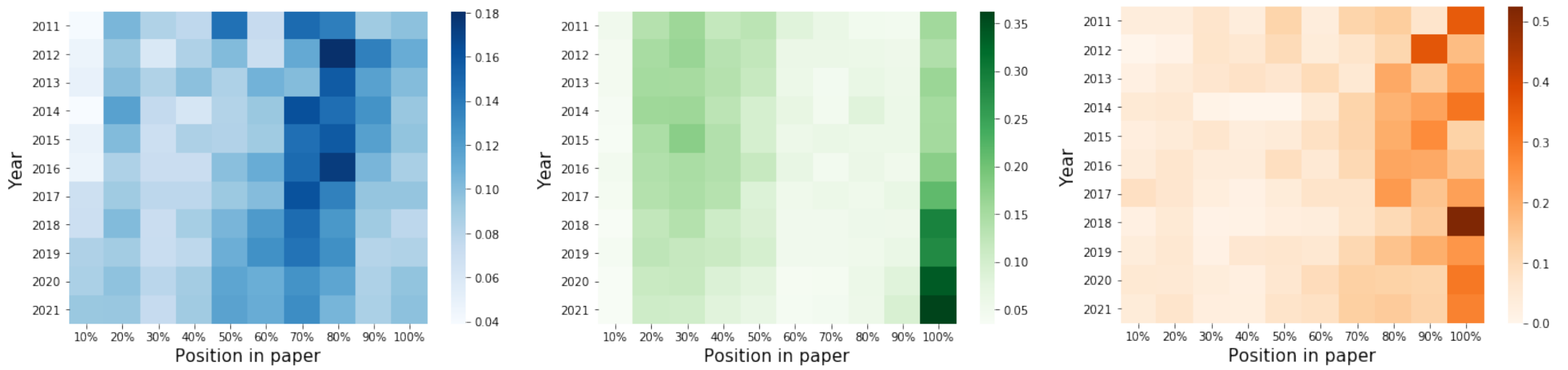

(a) CS (b) Physics (c) Math

**Fig. 18.** Likelihood of data links appearing in different positions of paper over years in different fields. Darker color indicate higher probability of finding material link in corresponding position for a given year. As can be observed, both physics and math are likely to attach data to the end of the paper, often in appendix or data availability section, while patterns in computer science is less obvious.

**Table 5. Logistic regression results on the relationship between link characteristics and the likelihood whether the link stay alive. More recent URLs are more likely be 200. CS URLs are more likely to be 200 than math. Physics URLs are less likely to be 200 than math. More popular URLs are more likely to be 200. Footnote URL is more likely to be 200.**

| Variable | Coefficient |
|---|---|
| Domain count (log) | 0.11 $***$ |
| URL count (log) | 0.24 $***$ |
| Paper citation count (log) | 0.016 $***$ |
| Whether the link is in footnote | 0.21 $***$ |
| The link appears in computer science (compared to math) | 0.22 $***$ |
| The link appears in physic (compared to math) | -0.25 $***$ |
| The link is a method link (compared to supplementary link) | -0.25 |
| The link is a data link (compared to supplementary link) | 0.016 $***$ |
| Age | -0.24 $***$ |
| Age squared | 0.0078 $***$ |

$^{***}p < 0.001$, $^{**}p < 0.01$, $^{*}p < 0.05$

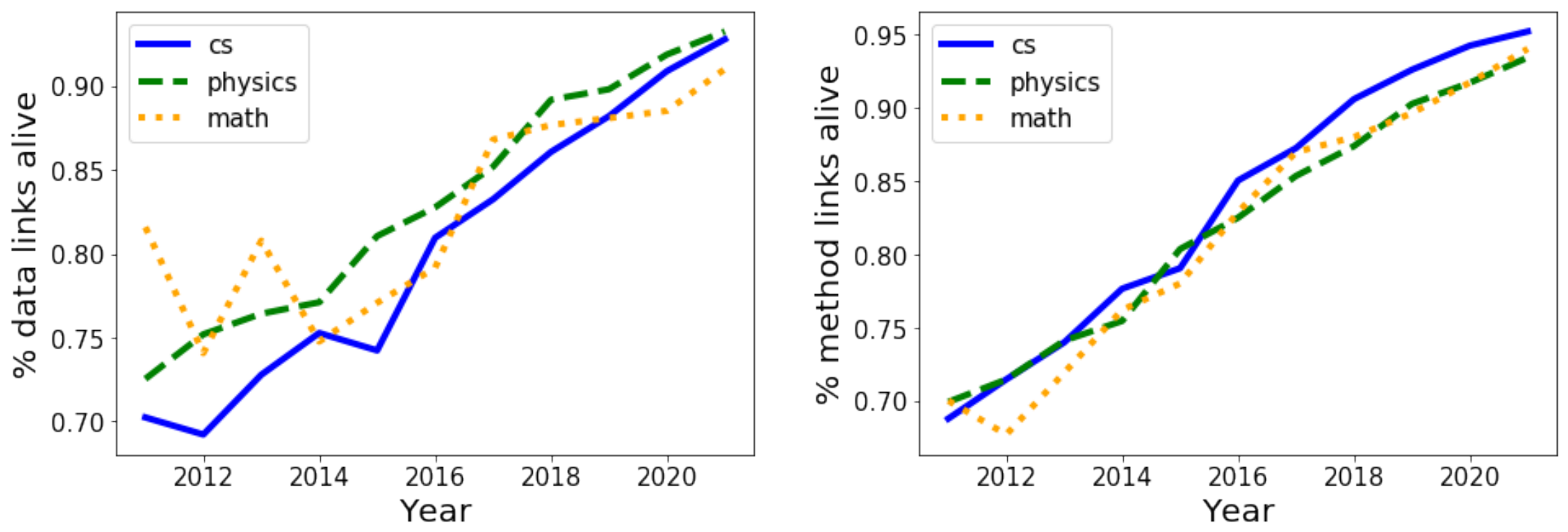

(a) Proportion of data links alive over time. (b) Proportion of method links alive over time.

**Fig. 19.** Proportion of data and method links alive across fields over time. Data and method links tend to be no longer retrievable over time.

**Table 6. Negative binomial regression results on the relationship between citation count and whether the paper contains (live) method and data links. We find haring data and methods links, and live versions of them, is associated with higher paper citation counts.**

| Variable | Coefficient |
|---|---|
| Contains live method link | 0.47 $***$ |
| Contains live data link | 0.27 $***$ |
| Contains dead method link | 0.08 |
| Contains dead data link | 0.14 |
| The link appears in math (compared to computer science) | -1.43 $***$ |
| The link appears in physic (compared to computer science) | -0.62 $***$ |
| Age | 0.69 $***$ |
| Age squared | -0.04 $***$ |

$^{***}p < 0.001$, $^{**}p < 0.01$, $^{*}p < 0.05$